\documentclass[
    ,final            
  ]
  {aipproc}

\layoutstyle{6x9}

\begin{document}

\title{The Light Scalar Mesons  $\mathbf{a_0/f_0(980)}$ at COSY-J\"ulich}


\author{M.~B\"uscher}{
  address={Institut f\"ur Kernphysik, 
    Forschungszentrum J\"ulich, 
    52425 J\"ulich, 
    Germany}
}



\begin{abstract}
  The light scalar mesons $a_0/f_0$(980) are being investigated at
  COSY-J\"ulich by detecting the strong decays into $K\bar K$ and
  $\pi\eta$/$\pi\pi$ as well as radiative decays into vector mesons.
  Selected results are discussed with emphasis on recent measurements
  at the ANKE and WASA spectrometers.
\end{abstract}

\maketitle

\section{Production of Light Scalar Mesons at COSY}

The COoler SYnchrotron COSY-J\"ulich provides proton and deuteron
beams --- phase-space cooled and polarized if desired --- with momenta
up to 3.7 GeV/c. It is thus well suited to produce the light scalars
$a_0/f_0$(980) since masses up to 1.1 (1.5, 1.03) GeV/c$^2$ can be
produced in $pp$ ($pd$, $dd$) collisions. Such hadronic interactions
offer a few particular advantages:

\begin{itemize}
\item Due to large production cross sections rare processes, like
  radiative decays into vector mesons $a_0/f_0\to \gamma
  V$~\cite{sraddec,Nagahiro:2008mn,radiativeproposal} or the
  isospin-violating $a_0$-$f_0$
  mixing~\cite{Hanhart:2003sk,ivproposal}, are expected to occur at
  reasonable count rates.

\item The isospin of the initial state and of the produced scalar
  meson can be selected. $pp\to dX^+$ or $pd\to tX^+$ reactions must
  lead to $a_0^+$ ($I{=}1$) production, a $pp\to ppX^0$, $pn\to dX^0$
  (using a deuterium target) or a $pd\to {}^3\mathrm{He}\, X^0$
  reaction can produce both the $a_0$ and the $f_0$, whereas the
  $dd\to{}^4\mathrm{He}\, X$ process is a filter for the $f_0$
  ($I{=}0$) resonance, because the initial deuterons and the $\alpha$
  particle in the final state both have isospin $I{=}0$.

\item For the $a_0/f_0\to K\bar K$ decays the maximum accessible
  excess energy $Q$ is rather small. Thus with a forward magnetic
  spectrometer like ANKE (see below), large acceptances and an
  unprecedented mass resolution $\delta_{m_{K\bar K}}$ can be reached.
  This is important to unravel effects induced by the opening of the
  $K^+K^-$ and $K^0 \bar K^0$ thresholds, which are separated by only
  8~MeV/c$^2$.
\end{itemize}

At the same time the following drawbacks should be mentioned:

\begin{itemize}
\item Also the cross sections for background processes like multi-pion
  production are large. 

\item The final states contain at least two baryons. Therefore, the
  scalar meson signal can be distorted by final-state interactions
  (FSI) between these baryons and/or between one or more baryons and
  the mesons from $a_0/f_0$ decays. This effect has, {\em e.g.\/} been
  seen for the $pp\to pp K^+K^-$ reaction~\cite{Maeda:2007cy} and also
  in $pd\to {}^3\mathrm{He}\,K^+K^-$ data~\cite{Grishina:2006cm}. On
  the other hand such experiments can be exploited for the
  investigation of the low energy $\bar{K}N$ and $\bar{K}A$
  interactions, see {\em e.g.\/} the analyses in
  Refs.~\cite{Sibirtsev:2004kk,Grishina:2005tu}.
\end{itemize}

$a_0/f_0$ production has been or will be studied at COSY in $pp$,
$pn$, $pd$ and $dd$ interactions for the strong decays into $K\bar K$
and $\pi\eta$/$\pi\pi$ as well as radiative decays $\gamma V$ into
vector mesons. While near-threshold decay channels with at least one
charged kaon can well be investigated with the magnetic spectrometer
ANKE (``Apparatus for the detection of Nucleonic and Kaon
Ejectiles''), the $\pi\eta$, $\pi\pi$ and $\gamma V$ final states will
be measured with WASA (``Wide Angle Shower Apparatus'') which is
available for measurements at COSY since 2007.

\subsection{ANKE spectrometer}
The magnetic ANKE spectrometer~\cite{Barsov:2001xj} consists of three
dipoles and detection systems for identification of charged particles
emitted under forward angles. For our measurements an H$_{2}$/D$_{2}$
cluster-jet target, which can provide areal densities of up to $5\cdot
10^{14}$ cm$^{-2}$s$^{-1}$, has been used. Together with $10^{11}$
particles in the COSY ring, this corresponds to luminosities up to a
few times $10^{31}$ cm$^{-2}$s$^{-1}$.

$K^{+}$-mesons are detected in a positive side detection
system~\cite{Buescher:2002zc}, using time-of-flight (TOF) measurement
between 23 scintillation start counters, which are placed near a side
exit window of the spectrometer magnet, and the range telescopes
system or a wall of scintillation counters. The momentum
reconstruction algorithm uses the track information provided by two
multiwire proportional chambers (MWPCs).  This information as well as
the kaon energy losses in the scintillators are used in order to
suppress background.  High momentum particles ($p$, $d$, $t$ or He)
produced in coincidence with the kaons are detected by a forward
detection system which consists of three MWPCs (used for momentum
reconstruction) and two layers of scintillation counters (particle
ID). As a selection criteria, the energy loss of the particles and
time difference between the hits in the side and forward systems are
used. $K^{-}$-mesons are observed in a negative side detection system
containing layers of scintillation counters and two MWPCs, which also
provide the possibility to use the time difference between negative
and positive detection systems, and $\Delta E$ techniques and to
reconstruct $K^{-}$ momenta~\cite{Hartmann:2007ks}.

\subsection{WASA spectrometer}
\label{wasa}
The 4$\pi$ detector facility WASA~\cite{Adam:2004ch} has been designed
for studies of production and decays of light mesons.  WASA makes use
of a hydrogen and deuterium pellet target. The pellet concept is
crucial to achieve a close to 4$\pi$ detection acceptance in this
internal target storage ring experiment.  The target system provides
small spheres of frozen hydrogen or deuterium and allows for high
luminosities of up to $10^{32}$ cm$^{-2}$s$^{-1}$.

The WASA detectors comprise a forward part for measurements of charged
target-recoil particles and scattered projectiles ($p$, $d$, $t$ or
He), and a central part for measurements of the scalar meson decay
products. The forward part consists of eleven planes of plastic
scintillators and of proportional counter drift tubes. The central
part comprises an electromagnetic calorimeter with $\sim 1000$ CsI(Na)
crystals surrounding a superconducting solenoid. Inside the solenoid a
cylindrical chamber with drift tubes and a plastic scintillator barrel
are placed.

\section{The $\mathbf{K\bar K}$ Final State}

Close-to-threshold data on kaon pair production in nucleon-nucleon
scattering, like in the reactions $pp \to d K^+\bar K^0$ or $pp \to
ppK^+K^-$, allow to study the $K\bar K$ and $\bar KN$ subsystems in
the final state. The strength of the $K\bar K$ interaction is of
relevance for a possible $K\bar K$ molecule interpretation of the
scalar resonances $a_0$(980) and $f_0$(980). Similarly, a better
understanding of the $\bar KN$ system is one prerequisite to infer the
nature of the $\Lambda$(1405).

\begin{figure}
  \includegraphics[width=0.8\textwidth]{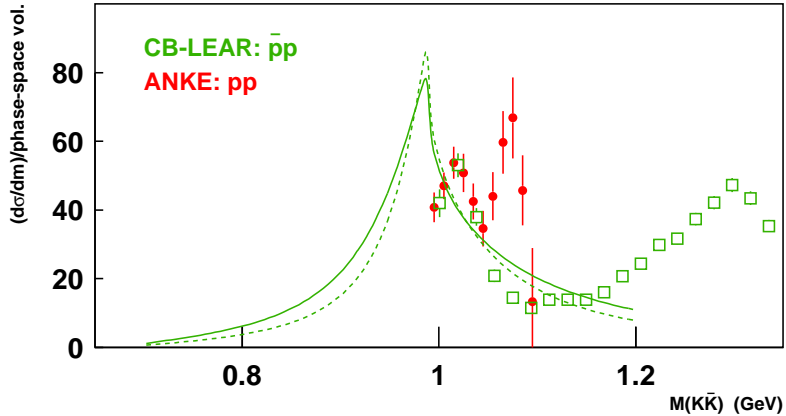}
  \caption{\label{fig:K+K0}Invariant $K^+\bar K^0$ mass distributions
    normalized to the phase-space volume. The Crystal-Barrel data
    (open squares) are from the $\bar p p\to K_L K^\pm \pi^\mp$
    reaction~\cite{Abele:1998qd}; the ANKE data (full circles) for
    $pp\to dK^+\bar K^0$ at $Q=105$ MeV~\cite{Dzyuba:2006bj}.
    The mass resolution of the ANKE data is $\delta_{m_{K\bar K}}
    =3\ldots 10$ MeV/c$^2$ (FWHM). The lines denote the Flatt\'e fits
    to Crystal-Barrel data from Refs.~\cite{Abele:1998qd} (solid)
    and \cite{Bugg:1994mg} (dashed).}
\end{figure}

The $pp \to d K^+\bar K^0$ reaction has been measured with ANKE at two
proton kinetic energies $T_p = 2.65$ GeV, and 2.83 GeV, corresponding
to excess energies of $Q = 48$ MeV and 105 MeV with respect to the
$dK^+\bar K^0$ threshold~\cite{Kleber:2003kx,Dzyuba:2006bj}.
Figure~\ref{fig:K+K0} shows the invariant $K^+\bar K^0$ mass
distribution (normalized to the phase-space volume) for the higher
beam energy. The data seem to indicate a two-peak structure which
clearly deviates from the expected Flatt\'e-like behaviour for the
$a_0^+$ resonance (indicated by the lines). While the enhancement at
higher $K\bar K$ masses probably is a reflection of the $\bar Kd$
FSI~\cite{ad_tbp}, there is no obvious interpretation of the lower
one. Interestingly, the same peak structure (although, due to the
lower mass resulution, only supported by the lowest-mass data point)
is also observed for data from $\bar pp$ annihilations. A possible
explanation of the unexpected shape is that interference effects by
chance lead to the same structure in both data sets.

\begin{figure}
  \includegraphics[width=0.35\textwidth,angle=270]{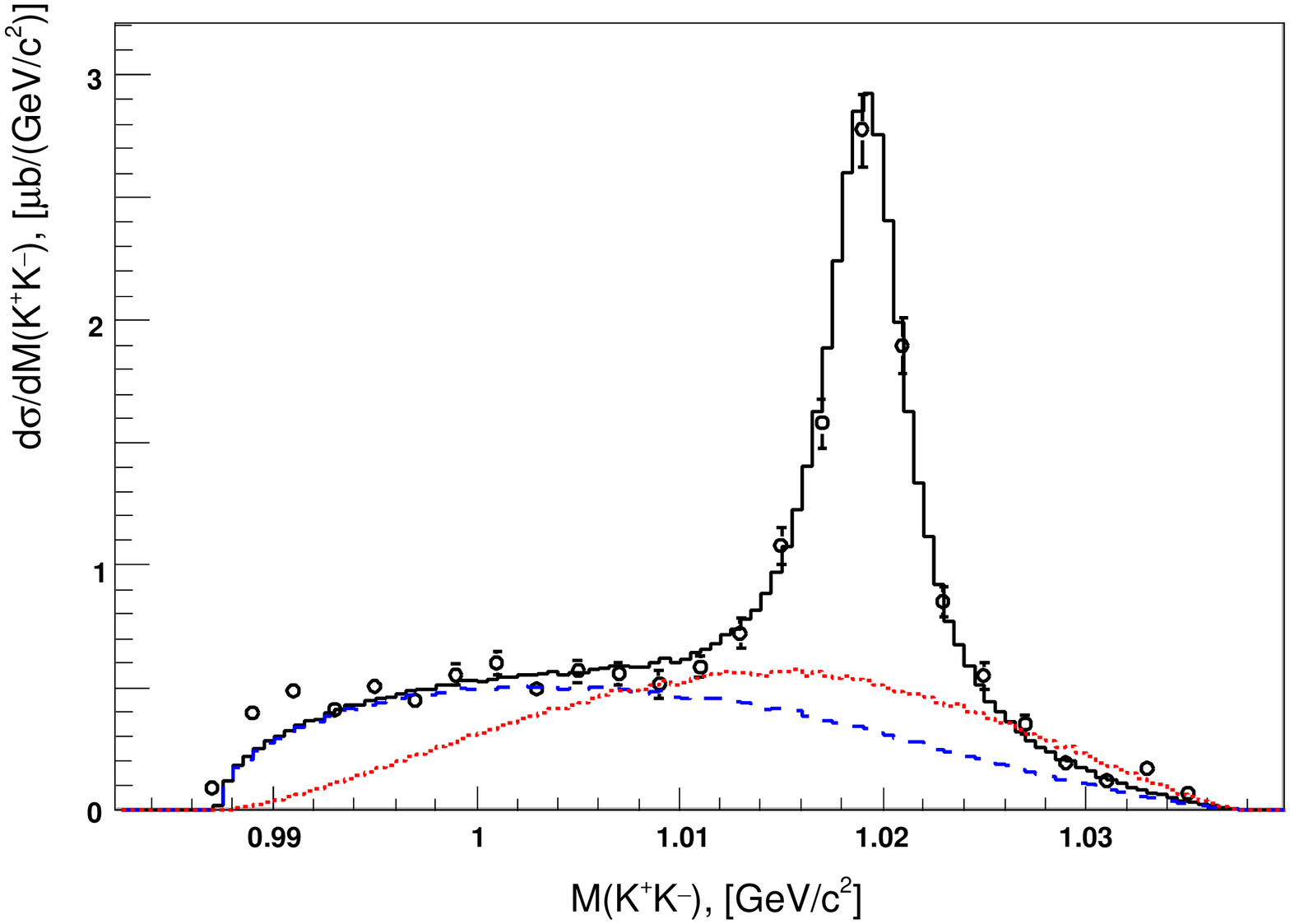}
  \includegraphics[width=0.35\textwidth,angle=270]{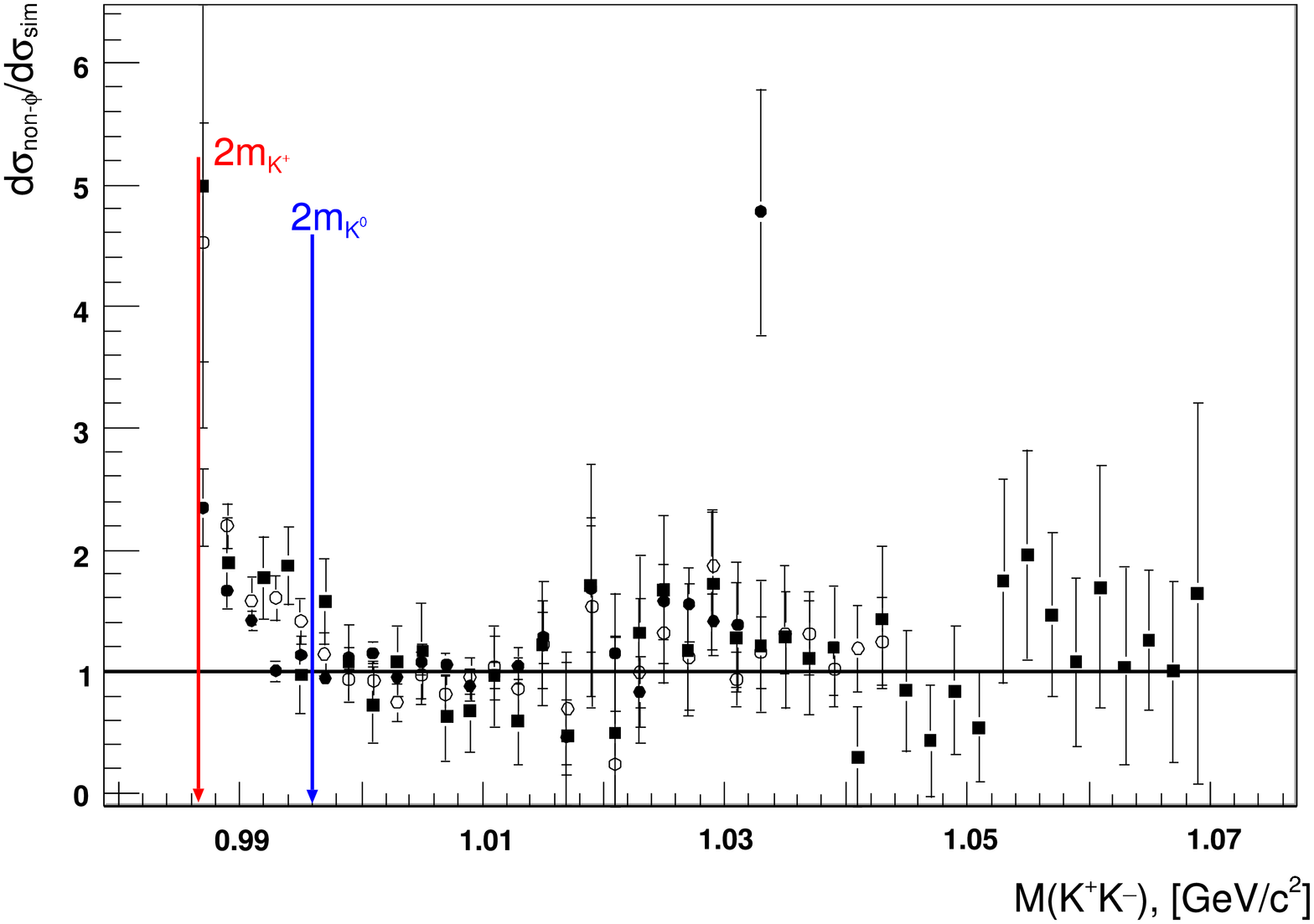}
  \caption{\label{fig:K+K-}Left: Differential cross section for the
    $pp \to ppK^+K^-$ reaction with respect to the $K^+K^-$ invariant
    mass at $T_p = 2.65$ GeV~\cite{Maeda:2007cy}. The solid (black)
    line shows the expected shape of the distribution from a
    Monte-Carlo simulation that takes into account the $\phi(1020)$,
    non-$\phi$ contributions as well as the $\bar Kp$ and $pp$ FSIs
    (best fit)~\cite{Maeda:2007cy}. For illustration, the dashed
    (blue) and dotted (red) lines show the shape of the non-$\phi$
    mass distribution with the $K^+K^-$ being in an S(P) wave.  Right:
    Data for all measured energies 2.65 GeV, 2.70 GeV, and 2.83 GeV
    normalized to the best-fit distributions. The mass resolution is
    $\delta_{m_{K\bar K}} = 2\ldots 3$ MeV/c$^2$ (FWHM).}
\end{figure}

The measurements of the $pp \to pp K^+K^-$ reaction were performed
at three energies of $T_p = 2.65$ GeV, 2.70 GeV and 2.83 GeV, {\em
  i.e.\/} at $Q$ values of 51 MeV, 67 MeV and 108 MeV with respect to
the $ppK^+K^-$ threshold. Figure~\ref{fig:K+K-} presents the invariant
$K^+K^-$ mass resolution for the lowest beam energy.  The lines show
the expected shape of the distribution from a Monte-Carlo simulation
that takes into account the ANKE acceptance~\cite{Maeda:2007cy}.  Three
data points at the lowest $K\bar K$ masses lie significantly above the
simulation. This behaviour is also visible for the other two beam
energies and in DISTO results on the same reaction at
$Q=110$~MeV~\cite{Balestra:2000ex} as well as in our data on the $pn
\to dK^+K^-$ reaction~\cite{Maeda:2006wv}. This effect is demonstrated
in the right part of Fig.~\ref{fig:K+K-} where the ANKE differential
cross section is normalized to the simulated spectra. At all three
beam energies a sigificant enhancement between the $K^+K^-$ and
$K^0\bar K^0$ thresholds is observed.

The mass scale of the low-mass variation in Fig.~\ref{fig:K+K-} is not
that of the widths of the scalar resonances, which are much larger. It
is tempting to suggest that this structure might be due to the opening
of the $K^0 \bar K^0$ channel at a mass of 0.995~GeV/c$^2$, which
induces some cusp structure that also changes the energy dependence of
the total cross section near threshold.  This would require a very
strong $K^+K^-\rightleftharpoons K^0 \bar K^0$ channel coupling, which
might be driven by the $a_0/f_0$ resonances.  Thus, although the $pp
\to pp K^+K^-$ reaction may not be ideal for investigating the
properties of scalar states, their indirect effects might still be
crucial.

\section{Searches for Isospin-Symmetry Violation}

An evident test of isospin symmetry is to scrutinize the $a_0^0$ and
$a_0^+$ mass distributions. So far, the best knowledge of the $a_0^0$
shape comes from a Flatt\'e fit to the $\pi^0\eta$ mass distribution
from $p\bar p$ annihilations, measured with Crystal Barrel at
LEAR~\cite{Bugg:1994mg}. The fit yields the coupling constants with
small statistical uncertainty, however, the $\pi^0\eta$ channel could,
in principle, be distorted by isospin-violating $a_0^0$-$f_0$ mixing
effects, {\em e.g.} distortions of a basically symmetric Breit-Wigner
type ($I{=}1,\, I_3{=}0$) $a_0$ mass distribution by admixtures of the
($I{=}0,\, I_3{=}0$) $f_0$ resonance.  This effect might be sizable if
the $\bar p p \to f_0 X$ cross section is significantly larger than
for the $a_0^0$ \cite{Amsler:1995bf,Amsler:1994pz}

\begin{figure}
  \includegraphics[width=0.7\textwidth]{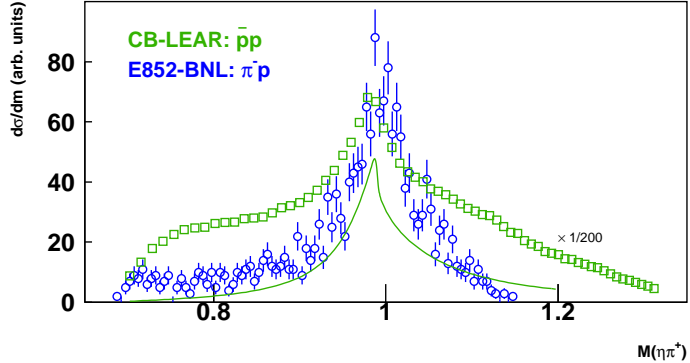}
  \caption{\label{fig:a0-vs.-a+}Invariant $\pi^0\eta$ mass
    distribution from $\bar p p\to \pi^0\eta\eta$ Crystal-Barrel data
    (green squares)~\cite{Amsler:1995bf} in comparison with that for
    the positive charge state $\pi^+\eta$ from pion-induced reactions
    (blue circles)~\cite{Teige:1996fi}. The line denotes a Flatt\'e
    fit from Ref.~\cite{Bugg:1994mg} to the Crystal-Barrel data.}
\end{figure}

It is thus desirable to obtain high statistics data for the $I_3{=}+1$
$a_0^+$ state, where mixing effects with the $f_0$ are strictly
forbidden. The best data so far on the $a_0^+\to \pi^+\eta$ decay
from an experiment at BNL on pion-induced
reactions~\cite{Teige:1996fi} are displayed in
Fig.~\ref{fig:a0-vs.-a+} together with the above mentioned
Crystal-Barrel data. The measured $a_0^+$ shape can as well be fitted
by a Flatt\'e distribution as by a Breit-Wigner (the latter
corresponds to the case of zero coupling of the $a_0^+$ to kaons).

It has been predicted that $a_0$-$f_0$ mixing can lead to a
comparatively large isospin violation in the reactions $pd \to
{}^3\mathrm{He}\, a_0^0({\to}\pi^0\eta)/f_0({\to}\pi\pi)$ and $pd \to
t\, a_0^+({\to}\pi^+\eta)$ close to the corresponding production
thresholds~\cite{Grishina:2001zj}. One may either test whether the
cross-section ratios for these reactions follow isospin relations, or
whether the $a_0^0$ and $a_0^+$ mass distributions exhibit different
shapes. Both require data with high statistical accuracy for the
strong scalar decay channels and backgound, {\em e.g.\/} from
non-resonant $\pi\eta$ production must be well understood. A first
test beam time took place in November 2007 at WASA and the data are
currently being analyzed.

\section{Outlook}

The $pd \to {}^3\mathrm{He}\, a_0^0/f_0$ and $pd \to t a_0^+$ data
that have been taken at WASA in November 2007 will also be used to
search for events from radiative decays into vector mesons $a_0/f_0\to
\gamma V$. According to our estimates~\cite{radiativeproposal} a few
thousand of such events could be measured in 10 weeks of beam time,
however, scheduling of such a long beam time at COSY has to await the
result of the above mentioned test measurement.

Another proposed measurement for WASA aims at data for the
isospin-violating $dd\to {}^4\mathrm{He}\, \pi^0\eta$
reaction~\cite{ivproposal}.  That cross section is expected to be
dominated by a primary reaction of the type $dd\to {}^4\mathrm{He}\,
f_0$, followed by an $f_0\to a_0$ conversion, and an
isospin-conserving $a_0\to \pi^0\eta$ decay.  In order to determine
the $f_0\to a_0$ mixing strength, the $dd\to {}^4\mathrm{He}\, f_0$
production cross section must be experimentally determined. For that
purpose, the $dd\to {}^4\mathrm{He}\, K^+K^-$ reaction has been
measured at ANKE.  A preliminary analysis yields a total cross section
in the range 50--100~pb.  If that cross section is dominated by the
$f_0\to K^+K^-$ decay, then isospin violation in the $dd\to
{}^4\mathrm{He}\, \pi^0\eta$ reaction should be measurable at WASA
within a few weeks of beam time.

\begin{theacknowledgments}
  This work has been supported by: Deutscher Akademischer
  Austausdienst, Deutsche Forschungsgemeinschaft, China Scholarship
  Council, COSY-FFE Program, Helmholtz-Gemeinschaft, Russian Academy
  of Science, Russian Fund for Basic Research, European Community.
  Contributions by the ANKE and WASA collaborations are greatfully
  acknowledged.
\end{theacknowledgments}

\end{document}